\documentclass[12pt]{iopart}

\usepackage{graphicx}
\usepackage{iopams}
\begin{document}

\title[High-efficiency single-photon Fock state production by transitionless quantum driving]{High-efficiency single-photon Fock state production by transitionless quantum driving}

\author{X. Shi$^{1,2}$ and L. F. Wei$^{2,3}$}

\address{Center of Quantum Information Technology, Chongqing Institute of Green and Intelligent Technology, Chinese Academy of Science, Chongqing, 400714, China}
\address{Quantum Optoelectronics Laboratory, School of Physics
and Technology, Southwest Jiaotong University, Chengdu 610031,
China}
\address{State Key Laboratory of Optoelectronic Materials and
Technologies, School of Physics Science and Engineering, Sun Yet-sen
University, Guangzhou 510275, China}
\ead{weilianfu@gmail.com}
\vspace{10pt}
\begin{indented}
\item[]November 2014
\end{indented}

\begin{abstract}
Single photon source is one of the key devices for optical quantum information processing. Differing from the usual stimulated Raman adiabatic passage to obtain single-photon radiation, here we propose an approach to produce optical Fock state on demand in the usual atom-cavity system by utilizing the technique of transitionless quantum driving. The present proposal effectively suppresses the unwanted but practically unavoidable nonadiabatic transitions in the previous adiabatic schemes. Therefore, the efficiency of Fock state production by the present technique could be significantly high, even in the presence of various atomic and cavity dissipations.
\end{abstract}

% Uncomment for PACS numbers
%
% Uncomment for keywords
\vspace{2pc}
\noindent{\it Keywords}: high efficiency, single photon, transitionless quantum driving
%
% Uncomment for Submitted to journal title message

\submitto{\LPL}
%
% Uncomment if a separate title page is required
\maketitle
%
% For two-column output uncomment the next line and choose [10pt] rather than [12pt] in the \documentclass declaration
%\ioptwocol

\noindent Due to the various applications in quantum information processing, the production of nonclassical states of light has been one of the most attractive topics both theoretically and experimentally~\cite{Beugnon,Maunz,Volz,Carmichael,Kuhn}. Physically, manipulating the atoms in a cavity by using the stimulated Raman adiabatic passage (STIRAP)~\cite{Hennrich}, single-photon Fock state emission could be achieved~\cite{Parkins}. However, owing to the practically-existing non-adiabatic transitions between the designed evolution passages~\cite{Parkins2}, the efficiency of Fock state production (FSP) is strongly limited~\cite{McKeever}.

To avoid the influence of non-adiabatic transitions on the efficiency and speed up the fast FSP, in this letter we propose an approach by using an alternative technique, named the transitionless quantum driving (TQD) or shortcut to adiabatic passage (SHAPE), to achieve the desirable FSP with high efficiency. The basic idea of this technique~\cite{Berry} is to find a driving $\hat{\mathbf{H}}(t)$ to evolve the system along a selected  instantaneous eigenstate $|\lambda_n\rangle$ of the initial Hamiltonian $\hat{H}_0$ exactly. During this driving any transition from this instantaneous eigenstate $|\lambda_n\rangle$ to the other ones \{$|\lambda_m\rangle, $ $m\neq n$\} is effectively suppressed.
Formally, the Hamiltonian for such a TQD can be expressed as $\hat{\mathbf{H}}(t)=\hat{H}_0(t)+\hat{H}_1(t)$~\cite{Berry} with
\begin{eqnarray}
\hat{H}_1(t)&=&i\hbar\sum_n|\partial_t\lambda_n\rangle
\langle\lambda_n|.\label{eq:1}
\end{eqnarray}
As $\hat{H}_1(t)$ is supplemented to $\hat{H}_0(t)$, the dynamics of the system can be restricted along the instantaneous eigenstate $|\lambda_n\rangle$ beyond the adiabatic limit. Therefore, this technique could be used to speed up the population passages for high-efficiency quantum state controls by introducing various classical pulses~\cite{Chen,Ibanez,Adolfo,Ruschhaupt,Chen2}. Here, with the full quantized atom-cavity interaction we discuss how to apply the TQD technique to achieve the high efficiency single-photon FSP with the usual double $\Lambda$-level atom interacting with a single-mode high-Q cavity.
Although the similar configuration had been used to implement the production of single photon by using the STIRAP technique (see, e.g.,~\cite{Brown,Gogyan1,Gogyan2}), the present proposal possesses a manifest advantage: the population transfer for the FSP could be implemented fast (as it is beyond the adiabatic limit) and deterministically (as it does not yield any unwanted leakage from the driven states). Therefore, the desirable FSP can be achieved with a significantly high efficiency.
\begin{figure}[htb]
\centerline{\includegraphics[angle=90,width=9cm]{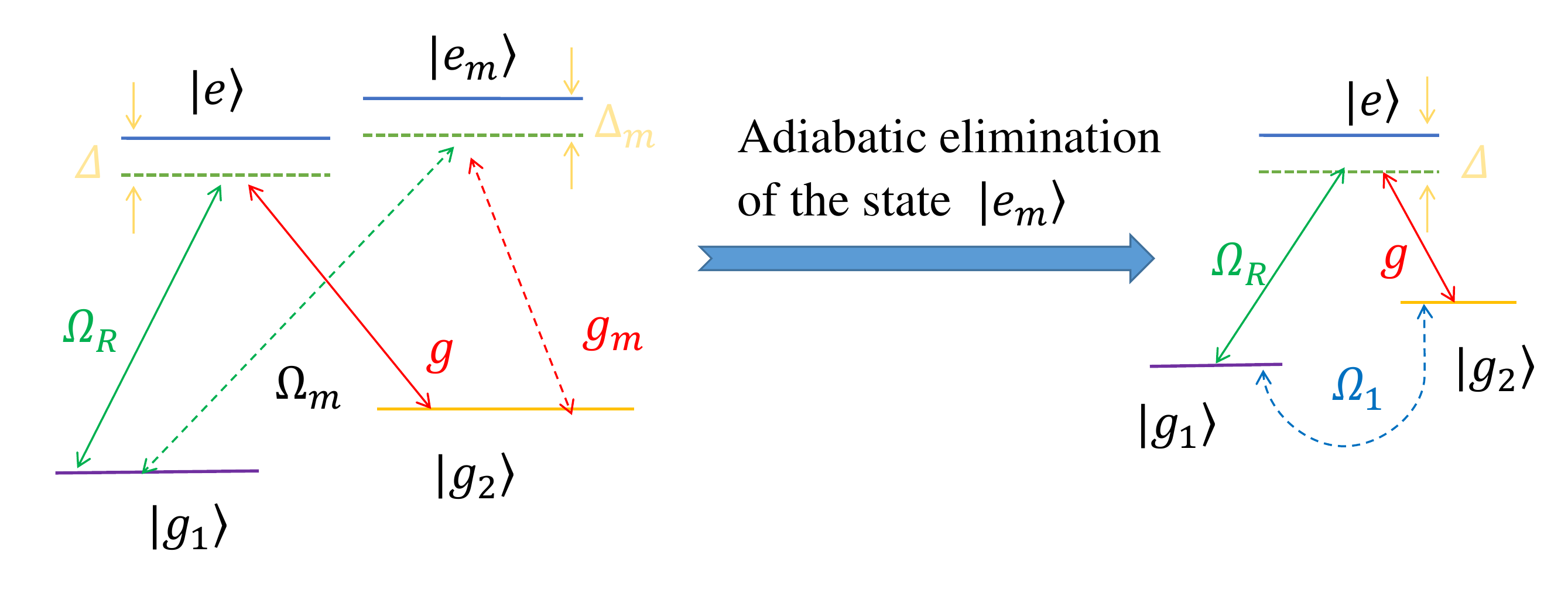}}
\caption{\label{fig:1}(color online). Effective three-level $\Lambda$ atomic configuration formed by adiabatically eliminating the auxiliary level $|e_m\rangle$.}
\end{figure}

The configuration of the atom-cavity system for our proposal is depicted in Fig.~\ref{fig:1}. The atom consists of two excited states $|e\rangle$ and $|e_m\rangle$, and two ground states $|g_1\rangle$ and $|g_2\rangle$. The transitions $|g_1\rangle\leftrightarrow|e\rangle$ and $|g_1\rangle\leftrightarrow|e_m\rangle$ are coupled by classical pumps $\Omega$ and $\Omega_m$, while transitions $|g_2\rangle\leftrightarrow|e\rangle$ and $|g_1\rangle\leftrightarrow|e_m\rangle$ is coupled to a cavity mode with coupling coefficients $g$ and $g_m$, respectively. Two detunings $\Delta$ and $\Delta_m$ are relative to the non-resonant transitions by the classical pumps and cavity couplings. In a rotating frame, defined by $\hat{U}=\exp{(-i\omega_{g_1}t)}|g_1\rangle\langle g_1|+\exp{[-i(\omega_{e}-\Delta) t]}|e\rangle\langle e|+\exp{(-i\omega_{g_2}t)}|g_2\rangle\langle g_2|+\exp{[-i(\omega_{e_m}-\Delta_m) t]}|e_m\rangle\langle e_m|$, the Hamiltonian of this system $\hat{H}_R$ reads ($\hbar=1$)~\cite{Parkins,Parkins2}
\begin{eqnarray}\label{eq:2}
\hat{H}_R(t)=\omega_c a^{\dagger} a+\Delta|e\rangle\langle e|+\Delta_m|e_m\rangle\langle e_m|
+\Omega_R S_1^{\dagger}\nonumber \\+g S_2^{\dagger}a
+\Omega_m F_1^{\dagger}+g_m F_2^{\dagger}a+\rm{H.c.}.
\end{eqnarray}
Here, $a$ ($a^{\dagger}$) is the annihilation (creation) operator of the cavity mode with the frequency $\omega_c$, $S_j^{\dagger}=|e\rangle\langle g_j|$ and $F_j^{\dagger}=|e_m\rangle\langle g_j|$ ($j=1,2$) are the raising operators of the atom. The detunings respect to the following equations,  $\Delta=\omega_{e}-\omega_{g_{1}}-\omega_L=\omega_{e}-\omega_{g_2}-\omega_c$, $\Delta_m=\omega_{e_m}-\omega_{g_{1}}-\omega_L=\omega_{e_m}-\omega_{g_2}-\omega_c$ (with $\omega_{g_1}$, $\omega_{e}$, $\omega_{e_m}$ and $\omega_{g_2}$ being the eigenfrequencies of the four atomic levels). It is seen that, the energy invariant subspace related to the Hamiltonian $\hat{H}_R(t)$ is formed by the
following dressed states: $|g_1,0\rangle$, $|e,0\rangle$, $|e_m,0\rangle$ and $|g_2,1\rangle$ (here we assume that the cavity mode is initially prepared in the vacuum state $|0\rangle$). Thus, the desirable single-photon FSP requires the system should be manipulated deterministically from the ground state $|g_1,0\rangle$ to the target state $|g_2,1\rangle$~\cite{Brown,Gogyan1,Gogyan2}. Although this progress had been achieved by the STIRAP technique with a $\Lambda$-level atom-cavity system, the efficiency of the desirable FSP is significantly limited due to the undesired non-adiabatic transitions. It is seen below that these transitions could be effectively suppressed by introducing the drivings related to the auxiliary excited state $|e_m\rangle$.

In principle, the Schr\"odinger equation for the Hamiltonian (2) can be rewritten as
\begin{eqnarray}\label{eq:3}
\left\{
  \begin{array}{c}
  i\dot{C}_1=\Omega_R C_2-i\Omega_m C_4\,\,\,\,\,\,\,\,\,\,\,\,\, \,\,\,\,\,\,\,\,\,\,\,\,\,\,\,\\
  i\dot{C}_2=\Omega_R C_1+\Delta C_2+g a C_3 \,\,\,\,\,\,\,\,\,\,\,\,\,\\
  i\dot{C}_3=g a^\dagger C_2+g_m a^\dagger C_4\,\,\,\,\,\,\,\,\,\,\,\,\,\,\,\,\,\,\,\, \,\,\,\,\\
  i\dot{C}_4=i\Omega_m C_1+g_m a C_3+\Delta_m C_4
  \end{array}\right.
\end{eqnarray}
in the atomic-level subspace $\{|g_1\rangle, |e\rangle, |g_2\rangle, |e_m\rangle\}$, with $|\Psi(t)=C_1|g_1\rangle+C_2|e\rangle+C_3|g_2\rangle+C_4|e_m\rangle$ and the "coefficients" $C_i\,(i=1,2,3,4)$. In the far-off-resonant case, the upper atomic state $|e_m\rangle$ can be effectively eliminated~\cite{Gogyan1}, i.e., $\dot{C}_4=0$. As a consequence, an effective Raman atom-photon coupling $\Omega_1=\Omega_m g_m/\Delta_m$ is delivered~\cite{Gogyan2}, and then the above double $\Lambda$-level atom-cavity system is reduced to an effective three-level $\Lambda$ one (see Fig.~\ref{fig:1}). Therefore, an effective Hamiltonian is delivered in the atomic space \{$|g_1\rangle$, $|e\rangle$, $|g_2\rangle$\}
\begin{eqnarray}\label{eq:4}
  \hat{H}_{eff}(t)&=& \left(
  \begin{array}{ccc}
  0 &\Omega_R &i\Omega_1 a  \\
  \Omega_R &\Delta &g a  \\
  -i\Omega_1 a^\dagger &g a^{\dagger} &0
  \end{array}\right).
\end{eqnarray}
During the derivation, all the Stark shifts of the relevant energy levels are neglected as they do not affect the following progress for the FSP. Formally, the above Hamiltonian (\ref{eq:4}) can be  written as $\hat{H}_{eff}(t)=\hat{H}_0(t)+\hat{H}_1'(t)$, wherein $\hat{H}_0(t)=\Delta|e\rangle\langle e|+\Omega_R S_1^{\dagger}+g S_2^{\dagger}a+\rm{H.c.}$ describes the usual progress of STIRAP in the effective three-level $\Lambda$ atom-cavity system, and $\hat{H}_1'(t)=i\Omega_1|g_1\rangle\langle g_2|a+\rm{H.c.}$ illustrates the effective Raman atom-photon coupling. Obviously, the auxiliary Hamiltonian $\hat{H}_1'(t)$ affects the efficiency of single photon generation.

In the subspace $\{|g_1,0\rangle, |e,0\rangle, |g_2,1\rangle\}$, one can easily check that the instantaneous eigenstates \{$|\lambda_n\rangle$\} of the Hamiltonian $\hat{H}_0(t)$ read, $|\lambda_0\rangle=\cos{\theta}|g_1,0\rangle+\sin{\theta}|g_2,1\rangle$, $|\lambda_+\rangle=\sin{\theta}\sin{\phi}|g_1,0\rangle+\cos{\phi}|e,0\rangle+
\cos{\theta}\sin{\phi}|g_2,1\rangle$, and $|\lambda_-\rangle=\sin{\theta}\cos{\phi}|g_1,0\rangle-\sin{\phi}|e,0\rangle+
\cos{\theta}\cos{\phi}|g_2,1\rangle$. Here, the mixing angles $\theta$ and $\phi$ are defined by $\tan{\theta}=\Omega_R /g $ and $\tan{2\phi}=2\Omega /\Delta$, whereas $\Omega=\sqrt{\Omega_R^2 +g^2 }$. In the previous STIRAP technique, the single-photon FSP is realized by the the dark state manipulation, i.e., the system evolves along the $|\lambda_0\rangle$-path via adiabatically adjusting the mixed angle $\theta$. However, when the adiabatic condition is not be satisfied exactly, the non-adiabatic transitions from the driven dark state to the bright states $|\lambda_\pm\rangle$ will take place\cite{Shi}. As a consequence, the above manipulation induces unwanted errors. In our proposal, once the effective Raman atom-photon coupling satisfy the condition $\Omega_1=[g\dot{\Omega}_R -\dot{g} \Omega_R ]/\Omega^2$ from Eq. (\ref{eq:1}), a shortcut to STIRAP is found and the undesired non-adiabatic transition can be suppressed~\cite{Berry}. Then, the population from $|g_1,0\rangle$ to $|g_2,1\rangle$, i.e., the expected single-photon emission can be achieved beyond the adiabatic limit.
\begin{figure}[htb]
\centerline{\includegraphics[width=9cm]{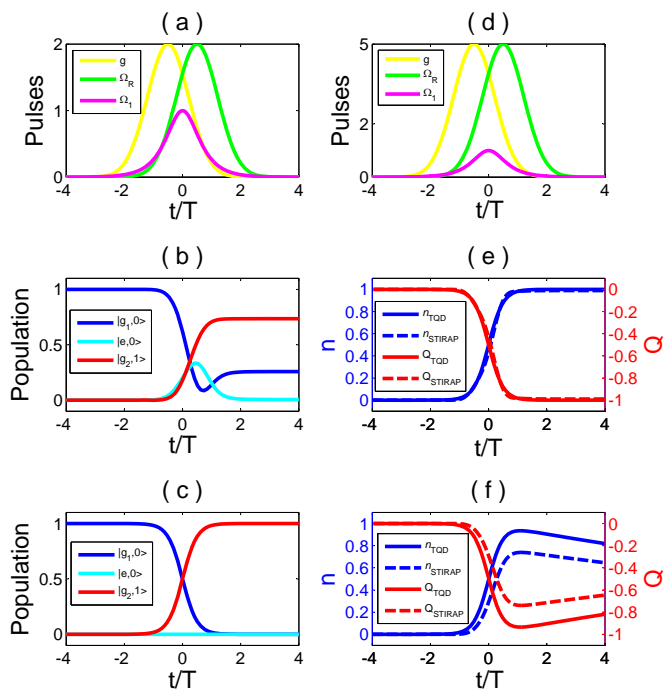}}
\caption{\label{fig:2} (color online). Comparison of single-photon FSPs by the STIRAP and TQD techniques. (a) Exciting pulses: $g$ and $\Omega_R$, are applied to realize the STIRAP; and the auxiliary driving $\Omega_1$ is additionally applied to implement the desired TQD with the parameter $\Omega_0 T=2$. Time-dependent populations during the dark state ($|\lambda_0\rangle$) evolution: (b) by the STIRAP, and (c) by the TQD, respectively.
(d) Exciting pulses and the auxiliary driving with the parameter $\Omega_0 T=5$.
Time-dependent mean cavity photon number $n$ and the Mandel $Q$ parameter during the TQD and STIRAP; without dissipation (e), and the presence of dissipation (f). Here, $\Gamma=5/T$ and $\kappa=0.05/T$.}
\end{figure}

Specifically, for the pulse sequence shown in Fig.~\ref{fig:2}(a) with the detuning $\Delta=1/T$ and the two drivings: $\Omega_R=\Omega_0\exp{[-(t-\tau_p)^2/T^2]}$ and $g=\Omega_0\exp{[-(t+\tau_s)^2/T^2]}$ (with $\tau_p=0.5T$ and $\tau_s=0.5T$), Fig.~\ref{fig:2}(b) (with $T=1\mu s$) shows that the usual STIRAP could not be perfectly realized, i.e., the final population of the target state $|g_2,1\rangle$ cannot reach unit (e.g., only about $73.5\%$).
However, when an auxiliary driving $\Omega_1$ is applied additionally, then the desired TQD is implemented and completely population passage from the initial state $|g_1,0\rangle$ to the target one $|g_2,1\rangle$ is achieved. This is numerically verified in Fig.~\ref{fig:2}(c), which indicates the single-photon Fock state of the cavity mode is generated perfectly. The above argument clearly shows that the TQD, rather than the usual STIRAP, provides a high efficient approach to deliver a single photon emission.

We now compared the robustness of the above STIRAP- and TQD-based single-photon FSP in the presence of the dissipations. In fact, atomic spontaneous emissions from excited state and photon absorption by the cavity mirrors are the main dissipation for the single photon generations~\cite{Fidio,Eleuch}. In the present of dissipation, the dynamics of the above effective $\Lambda$ atom-cavity system is alternatively described by the following master equation~\cite{Parkins,Parkins2}:
\begin{eqnarray}\label{eq:5}
\frac{\partial \hat{\rho}}{\partial t}=-i\left(
\hat{H}_{eff}^{\prime}\hat{\rho}-\rm{H.c.}\right)+
\kappa\hat{a}\hat{\rho}\hat{a}^\dagger
+\Gamma\sum_{j=1,2} S_j\hat{\rho}S_j^{\dagger},
\end{eqnarray}
with $\hat{H}_{eff}^{\prime}=\hat{H}_{eff}(t)
-i\Gamma|e\rangle\langle e|/2-i\kappa\hat{a}^\dagger\hat{a}/2$. Above, $\hat{\rho}$ is the reduced density operator, $\Gamma$ is the atomic spontaneous emission rate, and $\kappa$ the cavity dissipation.
By solving the above master equation, the time-dependent
mean photon number: $n=\langle \hat{a}^\dagger \hat{a}\rangle$ in the cavity, and the Mandel factor $Q=-1+[\langle (\hat{a}^\dagger \hat{a})^2\rangle-\langle \hat{a}^\dagger \hat{a}\rangle^2]/\langle \hat{a}^\dagger \hat{a}\rangle$ describing the single-photon quality can be obtained.
Specifically, we consider the pulse sequence schematized in Fig.~\ref{fig:2}(d). In the absence of dissipation, the ideal population passage for single-photon FSP can be implemented by both the designed STIRAP and TQD, see Fig.~\ref{fig:2}(e). However, Fig.~\ref{fig:2}(f) shows that in the presence of dissipation the quality of the single-photon emission generated by the TQD is obviously higher than that by the STIRAP.

Certainly, if the population transfer is strictly going alone the dark state evolution, i.e., the excited state $|e,0\rangle$ has been never populated, then the desired single-photon FSP will be realized perfectly. This requires the significantly large amplitude $\Omega_0$ of the drivings in the STIRAP.
However, in the TQD we do not care how strong the amplitude $\Omega_0$ but just apply a proper auxiliary driving $\Omega_1(t)$ to implement the exact dark state evolution.
During such a perfect dark state evolution the excited state $|e,0\rangle$ has never been populated. This is why the TQD is more robust than the STIRAP against the dissipation.
\begin{figure}[htb]
\centerline{\includegraphics[width=9cm]{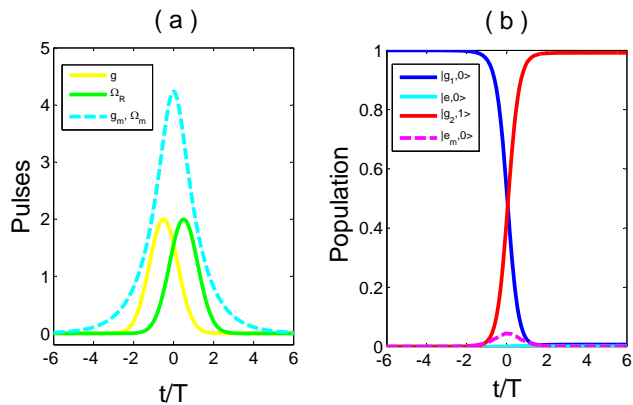}}
\caption{\label{fig:3} (color online). Single-photon FSP without adiabatically eliminating the level $|e_m\rangle$. (a) Exciting pulses: $g$, $\Omega_R$, $g_m$ and $\Omega_m$. Here, we consider the simplified case $g_m=\Omega_m$. (b) Time-dependent populations during the double $\Lambda$-level atom-cavity system.}
\end{figure}

To implement the technique of our TQD, the most crucial step is to realize the effective Raman atom-photon coupling between the states $|g_1,0\rangle$ and $|g_2,1\rangle$ by adiabatically eliminating the auxiliary state $|e_m\rangle$. For simplicity, we assume that the pulses $g_m$ and $\Omega_m$ are designed as the same shape, e.g., $g_m=\Omega_m=\alpha \exp{[-(t^2+\tau_s^2)/T^2]}/\beta$ with $\alpha^2=2\Delta_m/T$ and
\begin{eqnarray}\label{eq:6}
\beta=\sqrt{\exp{[-2(t+\tau_s)^2/T^2]}+\exp{[-2(t-\tau_p)^2/T^2]}}.
\end{eqnarray}
Then, the population dynamics of the double $\Lambda$-level atom-cavity system can be numerically solved and the relevant results are depicted in Fig.~\ref{fig:3} for, e.g., $\Delta_m T=18$. It is clear to see that the population dynamics of the state $|e_m\rangle$ is very small and completely population transfer from state $|g_1,0\rangle$ to $|g_1,1\rangle$ is robustly realized. This means that the auxiliary state $|e_m\rangle$ can be adiabatically eliminated really. Certainly, as $\Delta_m T\geq18$, the condition $\dot{C_4}=0$ still holds and thus we can still eliminate the state $|e_m\rangle$. Furthermore, Fig.~\ref{fig:3} gives a directly evidence that the efficiency of the single-photon FSP by the TQD is much higher than that by the STIRAP.

In summary, we have proposed a robust technique to produce single-photon Fock state on demand with a double $\Lambda$-level atom in a single-mode cavity. The auxiliary driving to implement the transitionless evolution is introduced by using an auxiliary level $|e_m\rangle$. By effectively eliminating such an level, the effective Raman atom-photon coupling makes the system evolve exactly along the relevant dark state path. In particular, we have also analyzed specifically the quality of the proposed TQD-based single-photon FSP by numerically solving the relevant master equation. The results showed that, the proposal is still robust in the presence of dissipation of the considered atom-cavity system. Hopefully, arbitrary multi-photon Fock states can also be produced by designing the relevant transitionless quantum drivings. Given the STIRAP technique have being taken the important role in atomic physics and quantum optics, the TQD-based FSPs proposed here should be implementable with the current cavity-QED experiments.
\ack
This work was supported in part by the National Science Foundation
grant Nos. 90921010, 11174373, U1330201, the National Fundamental Research
Program of China through Grant No. 2010CB923104, the 2013 Doctoral Innovation funds of Southwest Jiaotong University and the Fundamental Research Funds for the Central Universities.

\end{document}